\documentclass{article}

\usepackage[margin=3cm]{geometry}
\usepackage{amsmath,amssymb,amsfonts}
\usepackage[nocompress]{cite}
\usepackage{orcidlink}
\usepackage{hyperref}
\hypersetup{colorlinks=true,breaklinks=true,linkcolor=blue,filecolor=magenta,urlcolor=cyan}

\usepackage{authblk}
\newcommand{\keywords}[1]{\textbf{Keywords:} #1}

\begin{document}

\title{The Generalized Second Law and the Spatial Curvature Index}
\setlength{\affilsep}{3ex}

\author{Diego Pav\'{o}n\footnote{Email: diego.pavon@uab.es}\orcidlink{000-0001-9676-8716}}
\affil{Department of Physics, Autonomous University of Barcelona, \protect\\
08193, Bellaterra, Barcelona, Spain}


\maketitle

\begin{abstract}
\noindent By applying the generalized second law to the apparent horizon of a homogeneous and isotropic universe and imposing that the equation of state is no lower than $-1$, it is seen that universes with either flat or closed spatial sections are consistent with the joint consideration of the aforesaid law and the dominant energy condition but not always so universes with hyperbolic spatial sections.
\end{abstract}
\bigskip

\bigskip\bigskip\bigskip\bigskip

\section{Introduction}

Modern cosmology is mostly based on the cosmological principle; it  asserts that our universe is homogeneous and isotropic in the large scale average. By ``large scale" we mean scales larger than the biggest galaxy superclusters. Nearly a century ago, Robertson \cite{robertson} and Walker \cite{walker} derived the most general line element that satisfies the aforesaid principle and does not depend on any gravity theory. In comoving polar coordinates it takes the form
\begin{equation}
ds^{2} = -dt^{2} + a^{2}(t) \left[ \frac{dr^{2}}{1-k r^{2}} + r^{2} (d\theta^{2} + \sin^{2}\theta  d\phi^{2})\right],
\label{line-element}    
\end{equation}
where a(t) is the scale factor and $k$ the curvature index. The latter is either $-1$, $0$ or $+1$ depending on whether the spatial sections of our universe are hyperbolic (open), flat or closed, respectively. In the two first cases the volume of the said sections is unbounded while in the third case ($k = +1$) is finite.
\\

Tests of the cosmological principle have been carried out for many years using different observational techniques, on the isotropy side, see e.g. \cite{ellis1984, kitching2014, bengaly2019, guandalin2023, dinda2025}, and on the homogeneity side as well, see for instance \cite{clarkson2008, clarkson2010, zhang2011, camarena2021}. Nowadays, according to the general consensus, the aforesaid principle is, at large scales, a very good approximation to the real universe; see,  however, \cite{pietronero1998} and references therein.  
\\

As for the spatial curvature, evaluated by the dimensionless curvature parameter $\Omega_{k} = - k/(a^{2} H^{2})$, where $H =\dot{a}/a$ is the Hubble
factor, the overall agreement is that its absolute value is very small, if not nil (see, e.g. \cite{denissenya2018, divalentino2020, sunny2021, handley2021, bel2022, yang2023}). This is very reasonable since measurements of the present value of the spatial curvature, based on specific, reasonable, cosmological models, suggest that today its absolute value is of the order of $\mid 10^{-2}\mid $ or even less, i.e.  $\mid \Omega_{k0}\mid \simeq 0$\footnote{A sub-zero attached to any quantity refers to its present value.}. It is natural given the evolution of $\Omega_{k}$ with accelerated expansion. If  $\Omega_{k} >0$, $d{\Omega}_{k}/da < 0$ follows and vice-versa. Consequently,  its present value, whatever the initial one was at the beginning of the acceleration, is bound to be close to zero and it will more so at later times.  This implies that observationally it is very difficult to discard any of the three possible values of the curvature index. The aim of this short paper is precisely to eliminate one of them by means of theoretical arguments only. These just involve Einstein gravity, the  dominant energy condition (DEC) and the generalized second law (GSL). Therefore our reasoning does not rest on any particular model. As it turns out, universes with hyperbolic spatial sections, i.e. $k=-1$,  should be discarded. We present our arguments in the next section and discuss them in the last one. In our units $c = G = 1$.

\section{Combining the DEC and GSL}
Let us begin by considering an ever-expanding, homogeneous and isotropic universe that obeys Einstein gravity. Thus we can write,
\begin{equation}
3 H^{2} + 3 \frac{k}{a^{2}} = 8 \pi \rho     
\label{eq:friedmann}
\end{equation}
and
\begin{equation}
\dot{\rho} + 3H (\rho + p) = 0,
\label{eq:dotrho}
\end{equation}
where $\rho$ is the total energy density, i.e., radiation, matter (luminous and dark) and dark energy, and $p$ the total pressure. We define the cosmic  equation of state as $w =p/\rho$ which, obviously, depends on the scale factor.
\\

Now the question arises: Which are the limits, from above and from below of $w$? The first one is given by the requirement that the speed of sound,
$\sqrt{\partial p/\partial \rho}$,  does not exceed the speed of light. So, $w \leq 1$. Unfortunately a parallel reasoning cannot be applied to derive the lower limit since it can be negative, as it is the case when the universe is dominated by dark energy. Here we recall how straightforwardly the lower limit of $w$ can be determined with  the help of the DEC \cite{wald1984,eric2004}. In the present context the latter boils down to $\rho > 0$ alongside $\rho + p \geq 0$.

\noindent Thus, the lower bound
\begin{equation}
w \geq -1
\label{eq;wlower}
\end{equation}
on the cosmic equation of state readily emerges.
\\   \

Next,  we will see that the GSL alone does not restrict the sign of $k$. We proceed as follows.  Equations (\ref{eq:friedmann}) and (\ref{eq:dotrho}) taken together produce
\begin{equation}
w = - \frac{1}{1 - \Omega_{k}} \left[ 1 + \frac{2}{3} \frac{\dot{H}}{H^{2}} + \frac{1}{3} \Omega_{k}\right],
\label{eq:wtotal}
\end{equation}
where $\mid \Omega_{k}\mid < 1$.
Note that this expression is valid regardless the various matter-energy species interact with each other or not. Clearly $w$ will be as low as $-1$ whenever $H$ remains constant  and $\Omega_{k} = 0$.
\\

Now, we may ask:  Will $w$ also have a lower limit when $k \neq 0$? The answer is yes provided that the GSL for cosmic horizons \cite{Davies1987,dpc,dpc1,dpc2} holds. 

\noindent To see this recalls that  the deceleration parameter is given by $q =-1-(\dot{H}/H^{2})$, Then we can be express (\ref{eq:wtotal}) in terms of the said parameter
\begin{equation}
w = - \frac{1 - 2 q + \Omega_{k}}{3 (1 - \Omega_{k})}.
\label{eq:wq}
\end{equation}

\noindent The GSL implies that the area of the apparent horizon\footnote{The boundary hyper-surface of the space-time anti-trapped region.},  ${\cal {A}} = 4 \pi/[H^{2} + k a^{-2}]$, never decreases in a expanding universe \cite{{bak-rey2000}}. That is to say, ${\cal {A}'} = - ({\cal {A}}^{2}/2 \pi) (H H' - k a^{-3}) \geq 0$, where the prime means $d/da$. This readily translates into $1+ q \geq \Omega_{k}$ \cite{npd2023a,npd2023b}. Using this in eq. (\ref{eq:wq}) it follows $w \geq \zeta$, where
\begin{equation}
\zeta =   \frac{\textstyle{1\over{3}} \Omega_{k} - 1}{1 - \Omega_{k}}
\label{eq:zeta}
\end{equation}
is the lower limit of the cosmic equation of state, $w$, valid at any redshift. Notice that for flat universes with De Sitter expansion $\zeta$ reduces to $-1$, as it should. Notice that for $\Omega_{k} > 0$ this  limit gets less than $-1$ (while $w \geq -1$) and diverges for $\Omega_{k} \rightarrow 1$. This leads  us to suspect that $\Omega_{k}$ may not be positive even though  it is compatible with the GSL and DEC when taken separately.  This prompt us to combine one with the other to see whether there is nothing wrong with it, nonetheless. 
\\ 

To this end we proceed as follows. The simplest linear combination of $w \geq -1$ and $w \geq \zeta$ is, 
\begin{equation}
(1+ \lambda) w \geq -1 + \lambda \left(\frac{\textstyle{1\over{3}} \Omega_{k} - 1 }{1 -\Omega_{k}}\right)
\label{eq:1+lambda}
\end{equation}
with $\lambda$ a continuous parameter satisfying $\lambda > -1$.  The latter inequality follows from a careful inspection of (\ref{eq:1+lambda}) and bearing in mind that $-1 < \Omega_{k} < 1$.
\\

Introducing the expression (\ref{eq:wq}) for $w$  in  (\ref{eq:1+lambda}),  we obtain after a short algebra
\begin{equation}
1 + q \geq  g(\lambda) \, \Omega_{k},
\label{eq:shortalgebra}
\end{equation}
where 
\begin{equation}
g(\lambda) = \frac{3}{2} \left[\frac{1+ \frac{\lambda}{3}}{1 + \lambda} + \frac{1}{3}\right]
\label{eq:g-lambda}    
\end{equation}
is an ever decreasing function.
\\

First, we note that the deceleration parameter lies in the range $-1 + \Omega_{k} \leq q \leq 1$\footnote{The upper bound on $q$ corresponds at early times when the universe was dominated by massless radiation. The lower bound, corresponds at late times when the universe is accelerating and one uses the DEC, under the form $\dot{H} \leq k/a^{2}$, alongside the expression for $\Omega_{k}$.}. On the other hand, the above constrain on $\lambda$ implies  $g(\lambda) \geq 1$, which fully agrees with the GSL, $1+q \geq \Omega_{k}$ (which is satisfied whatever the sign of $\Omega_{k}$). Hence, the physically meaningful range of the parameter  $\lambda$ is $-1 < \lambda < \infty$. (The other range, $-\infty < \lambda < -1$, is not to be considered because there $g(\lambda) < 1$). Note, by passing, that $g(\lambda)$ is not bounded from above.
\\ 

Clearly, for $\Omega_{k} < 0$ and $\Omega_{k} = 0$ (corresponding to $k = +1$ and $k = 0$, respectively) eq. (\ref{eq:shortalgebra}) is satisfied by any pair  of $q$ 
and $\lambda$ values lying in their respective intervals just mentioned. By contrast, if $\Omega_{k} > 0$ ($k = -1)$ the said equation is violated since its left-hand side is bounded from above ($1 + q \leq 2$), while  its  right-hand side is unbounded from above ($1 < g(\lambda) < \infty$). That is to say, for whatever $1+q$ value there are infinitely many $\lambda$ values (let be $\lambda_{0}$ any of them) such that $1+q < g(\lambda_{0}) \Omega_{k}$, thus violating eq. (\ref{eq:shortalgebra}).   
\\

Therefore, we may conclude that homogeneous and isotropic universes with hyperbolic spatial sections are not consistent with the DEC and GSL combined together.  
\noindent We wish to emphasize that this result rests only on Einstein gravity, the DEC (that implies $w > -1$) and the GSL (which leads to $w > \zeta$) and it is  independent on any particular cosmological model.
\\

The following example helps illustrate  this. Assume that from a late time ($a \gg 1$) on the universe is dominated  by pressureless matter and the cosmological constant $\Lambda$.

\noindent Then, the scale factor will obey the pair of Einstein equations
\begin{equation}
3H^{2} + 3 \frac{k}{a^{2}} = 8 \pi \rho_{m} + \Lambda   \qquad \quad (\rho_{m} = \rho_{m0} \, a^{-3})
\label{eq:m+L}
\end{equation}
and
\begin{equation}
3 \frac{\ddot{a}}{a} = -4 \pi \rho_{m} + \Lambda.
\label{eq:acceleration}    
\end{equation}
From these two we get, 
\begin{equation}
\frac{\ddot{a}}{a} \, - \, \left( \frac{\dot{a}}{a}\right)^{2} \,
- \frac{k}{a^{2}} = -4 \pi  \rho_{m}.
\label{eq:scalefactor}
\end{equation}

\noindent Since $ a \gg 1$ eq. (\ref{eq:scalefactor}) can be approximated, to any desired accuracy, by
\begin{equation}
a\, \ddot{a} \, - \, (\dot{a})^{2} \, - \, k = 0
\label{eq:scalefactor2}
\end{equation}
whose solutions, depending on the value of $k$, are
\begin{equation}
a(t)=\left\{
\begin{array}{lll}
H_{\star}^{-1}\, \text{cosh}(H_{\star} t)\,, & \quad & k = +1\\
a_{\star} \, \exp(H_{\star} t)\,, & \quad & k = 0\\
H_{\star}^{-1}\, \text{sinh}(H_{\star} t)\,, & \quad & k = -1
\end{array}
\right.
\label{desitter}
\end{equation}
where $H_{\star}\equiv \sqrt{{\Lambda}/{3}}$. The corresponding
expressions for the deceleration parameter, $q = -1
-(\dot{H}/H^{2})$, are,
\begin{equation}
q(t)=\left\{
\begin{array}{lll}
-1 \, - \, \frac{1}{\sinh^{2}(H_{\star} t)} \,, & \quad & k = +1\\
- 1 \,, & \quad & k = 0\\
- 1 \, + \frac{1}{\cosh^{2}(H_{\star} t)}\,, & \quad & k = -1 \, .
\end{array}
\right. \label{qdesitter}
\end{equation}
In its turn, the corresponding expressions for $\Omega_{k}$ are
\begin{equation}
\Omega_{k} =\left\{
\begin{array}{lll}
- \frac{1}{\sinh^{2}(H_{\star} t)} \,, & \quad & k = +1\\
0 \,, & \quad & k = 0\\
\frac{1}{\cosh^{2}(H_{\star} t)}\,, & \quad & k = -1 \, .
\end{array}
\right. 
\label{omegakdesitter}
\end{equation}
That is to say, in all three cases $1+q$ equals the corresponding $\Omega_{k}$ at any time thereby it is not possible
to discard any of the three possible $k$ values just by considering the relationship $1+q \geq \Omega_{k}$.
\\

However, things fare differently when one resorts to eq. (\ref{eq:shortalgebra}).
Using the expressions for $q$ and $\Omega_{k}$ of above, it is immediately seen that for $k = +1$ as well as for $k = 0$ the said equation is trivially satisfied (recall again that $g(\lambda) > 1$) while it is violated for $k = -1$, as expected. So, this simple example shows that the possibility $\Omega_{k} > 0$ (i.e. $k= -1$) gets  excluded on general grounds.   
\\

This is fully consistent with recent findings by Di Valentino et al. \cite{divalentino2020} suggesting that the Planck power spectrum favors a closed universe at 3-4 standard deviations ($-0.007 > \Omega_{k} > -0.095$) at 99\% C.L. See also \cite{handley2021} and figure 17 in \cite{yang2023}. However, even more recently Chen and Zaldarriaga \cite{Chen-Zaldarriaga2025} have proposed a model that reconciles the last published data on BAO with the CMB data and shows a $2\sigma$ preference  for an  open ($\Omega_{k} >0$) universe. Nevertheless, this does not rule out any of the other two curvature possibilities; but it manifests that nowadays, from the observational side, the sign of the spatial curvature is far from settled.

At this point one may wonder if some physical intuition can be adduced about the result that hyperbolic sections get excluded just using the GSL ad the energy conditions. We think so. In the first place, the volume of the spatial sections is unbounded. This implies an infinite amount of matter and energy in the universe. Clearly, this is counterintuitive and hard to accept both from the physical and  philosophical sides. In the second place, given the range of variability of the deceleration parameter, the fulfillment of the expression $1+q \geq \Omega_{k}$ of the GSL is automatically ensured for models featuring either $\Omega_{k} = 0$ or $\Omega_{k} < 0$ but not so for models with $\Omega_{k} > 0$, when the universe is accelerating its expansion ($q <0$). See also the paragraph around eq. (\ref{eq:zeta}). Hence, in principle, one may devise cosmological models with $\Omega_{k} > 0$ that violate the GSL and still comply with the DEC.  Therefore it is only natural to inquire whether open universes are compatible with the GSL and DEC taken simultaneously. Specifically, the hyperbolic FLRW model of late time, dominated by curvature and a cosmological constant (eq. (14) and the case $k= -1$ in equations in (15), (16) and (17)) is a clear example of incompatibility.

\section{Discussion}
By combining the dominant energy condition with the generalized second law in a homogeneous, isotropic and ever-expanding universe we have found that curvature parameter $\Omega_{k}$ can be either negative or nil but not positive; i.e., that the spatial sections of this universe cannot be hyperbolic. 
\\

It may be counter-argued that there are cosmological models with hyperbolic spatial sections that, one the one hand obey the GSL and, independently,  on the other hand, their equation of state always remains above $-1$. However, these models are unrealistic since they overlook the fact that the GSL and the DEC are not really independent (they should not be taken separately) as they affect each other since $q$ enters the expression of $w$, eq. (\ref{eq:wq}), and it is bounded by the GSL. As a consequence these models ignore the restriction $w \geq \zeta$  (with $\zeta$ given by eq. (\ref{eq:zeta})), and that the lower bound of $q$ is not $-1$ but $-1+\Omega_{k}$.
\\

Before closing it seems worthwhile mentioning that we have not considered phantom behavior because phantom fields, aside of violating the DEC, are afflicted by severe classical \cite{Dabrowski2015} and quantum \cite{cline2004} instabilities.    
\\

Altogether, if future cosmic observations finally  demonstrate that the spatial sections of our universe are hyperbolic (modulo the cosmological principle holds true), then at least, Einstein gravity or the DEC or the GSL fail.

\end{document}